\documentclass{iaus}
\usepackage{graphicx}

\def\Msun{M_\odot}

\def\go{\mathrel{\raise.3ex\hbox{$>$}\mkern-14mu
             \lower0.6ex\hbox{$\sim$}}}
\def\lo{\mathrel{\raise.3ex\hbox{$<$}\mkern-14mu
\lower0.6ex\hbox{$\sim$}}}

\title[Multiple Weak Deflections]{The Effects of Multiple Weak Deflections
in Galaxy--Galaxy Lensing}

\author[T. G. Brainerd]%
{Tereasa G. Brainerd}

\affiliation{Boston University, Institute for Astrophysical Research, \\
725 Commonwealth Ave., Boston, MA 02215 \\ brainerd@bu.edu}

\pubyear{2004}
\volume{225}
\pagerange{1--12}
\date{?? and in revised form ??}
\jname{Impact of Gravitational Lensing on Cosmology}
\editors{Mellier, Y. \& Meylan,G . eds.}
\begin{document}

\maketitle

\begin{abstract}
The Hubble Deep Field (North) and the flanking fields
are used investigate the occurrence of
multiple weak lensing deflections along the line of sight in relatively
deep imaging data ($z_{\rm lens} \sim 0.6$, $z_{\rm source} \sim 1.2$).
Ray tracing simulations of galaxy--galaxy lensing in the HDF-North show
that proper inclusion of multiple weak deflections is important for
a correct prediction of the net shear for most sources, and for 
a given source redshift the number
of multiple weak deflections is largely
insensitive to the cosmography.  
The effects of multiple weak deflections on the magnitude of the weak
lensing signal are, of course, strong functions of the adopted halo parameters.
Independent of the halo parameters,  however, the closest lens to a source
(in projection on the sky) is not the strongest lens in the case of more than
50\% of the sources which acquire
a net shear of $\gamma \lo 0.01$. 
In addition,
multiple weak deflections result
in a tangential shear about the lens centers that is greater than the
tangential shear that would occur if source galaxies were lensed solely
by the closest lens.  Further, multiple weak
deflections give rise to correlated image ellipticities and account
for a substantial amount of the total cosmic shear signal on small
angular scales in $\Lambda$CDM and open CDM models.
\end{abstract}

\section{Introduction}

It is now generally agreed that it will be possible in the near future
to obtain precision cosmological results via weak lensing measurements.
However, this statement is often interpreted to mean merely that with
appropriate due diligence on the observational end, precision constraints
will result.  The focus of this paper is to remind the reader
that highly accurate theory is equally important to the goal of placing
strict constraints on cosmology.  Simon White expressed this very nicely
in his review talk on numerical simulations when he said simply ``Precision 
cosmology will require precision simulations.''   

This paper attempts to demonstrate the truth of Simon's statement by
studying the effects of multiple weak lensing deflections on both the
galaxy--galaxy lensing signal and the cosmic shear signal.  
Galaxy--galaxy lensing has come a long way since the 1995 IAU
Symposium on gravitational lensing, where it appeared on the programme
in the ``Emerging Techniques'' session (Brainerd, Blandford \&
Smail 1996a).  Due to
a great deal of work by a number of different groups, galaxy--galaxy
lensing has not only been detected with impressively high statistical
significance, but it is also being used to place strong constraints on
the nature of dark matter halos and on the bias between light
and mass in the universe (see, e.g., 
Fischer et al.\ 2000;
McKay et al.\ 2001;
Hoekstra, Yee, \& Gladders 2004;
Kleinheinrich et al.\ 2003; Kleinheinrich 2004;
Natarajan et al.\ 220;
Seljak 2004)

Simulations
of galaxy--galaxy lensing for a moderately deep imaging survey
($I_{\rm lim} \sim 23$) performed by Brainerd, Blandford \& Smail (1996b),
hereafter BBS, showed that most of the galaxies with
magnitudes in the range
$22 \lo I \lo 23$ would, in fact,
 have been lensed at a comparable level by two or
more foreground galaxies (e.g., \S3.5 of BBS).  That is, multiple deflections
are expected to occur quite frequently in deep data sets and, in particular,
when comparing predictions for cosmic shear from a simulation to observations
of cosmic shear on small angular scales, 
it is important that the simulation reproduce faithfully
all of the weak galaxy lenses along the line of sight.  In other words, a
fair comparison between observations and theory on the scales for which
galaxy--galaxy lensing is important depends crucially on the ability of
simulations to follow the growth of 
the non--linear power spectrum accurately.  Even
with current codes and computer architecture, this is still somewhat
challenging and appropriate care must be taken
(see, e.g., Simon White's review in these proceedings).

The Hubble Deep Field (North) and the flanking fields have
been the subject of a deep redshift
survey (Cohen et al.\ 2000) as well as an extensive multicolor photometric
investigation (Hogg et al.\ 2000).  As a result, both the redshifts, $z$,
and the rest--frame blue luminosities, $L_B$, of $\sim 600$ galaxies in this
region of space are known (Cohen 2002).  Therefore, it is possible to make
quite a detailed theoretical prediction for the weak galaxy--galaxy lensing
shear field in the region of the HDF--North and, specifically,
for the probability and effects of
multiple deflections due to galaxy--galaxy lensing.

\section{Halo Properties and Fiducial Model}

For simplicity, the approach of BBS in their galaxy--galaxy
lensing simulations was adopted and the physical properties of the dark
matter halos around the galaxies in the HDF--North and the flanking fields
were scaled in terms of the characteristic
properties associated with the halos of $L^\ast$ galaxies.
The velocity dispersion of
an $L^\ast$ galaxy halo is given by $\sigma_v^\ast$ and it was assumed
that a Tully--Fisher or Faber--Jackson type of relation held
for each of the galaxies. Therefore, 
\begin{equation}
\frac{\sigma_v}{\sigma_v^\ast} = \left( \frac{L_B}{L_B^\ast} \right)^{1/4},
\end{equation}
where $\sigma_v$ is the velocity dispersion of a halo in which a galaxy with
luminosity $L_B$ resides.
The density profile of the galaxy halos was taken to be
\begin{equation}
\rho(r) = \frac{\sigma_v^2 s^2}{2\pi G r^2 \left( r^2 + s^2 \right) },
\end{equation}
where $G$ is Newton's constant and $s$ is a characteristic halo radius.
Further, it was assumed that the 
mass--to--light ratio of a galaxy was constant independent of its luminosity
and, therefore, the radii of the halos of galaxies with $L_B \ne L_B^\ast$
scale with the radii of the halos of
$L_B^\ast$ galaxies according to:
\begin{equation}
\frac{s}{s^\ast} = \left( \frac{L_B}{L_B^\ast} \right)^{1/2}.
\end{equation}
The total mass of the halo of an $L^\ast$ lens galaxy is finite and given by:
\begin{equation}
M^\ast = \frac{\pi s^\ast (\sigma_v^\ast)^2}{G}.
\end{equation}

Having made these assumptions, it is then possible to make predictions for
the galaxy--galaxy lensing shear field within the region of the
HDF--North that would
be generated by the
galaxies in Cohen (2002).  The lens galaxies include not
only galaxies in the HDF--North, but also galaxies in the flanking fields.
Because some of the galaxies in the flanking fields have quite substantial
masses, 
the weak lensing effects of these galaxies
have the potential to affect the shear field inside the much smaller region of
the HDF--North. 
Therefore, the flanking field galaxies were included
in all of the calculations for the weak shear field inside the HDF--North
itself.  Also, in order to have a consistent limiting magnitude for the
galaxy lenses, only those galaxies in Cohen et al.\ (2000) with 
$R \le 23$ were used in the calculations (i.e., the completeness limit
of the full redshift survey is deeper in the HDF--North than it is in the flanking
fields).

\begin{figure}
\centerline{
\scalebox{0.65}{%
\includegraphics{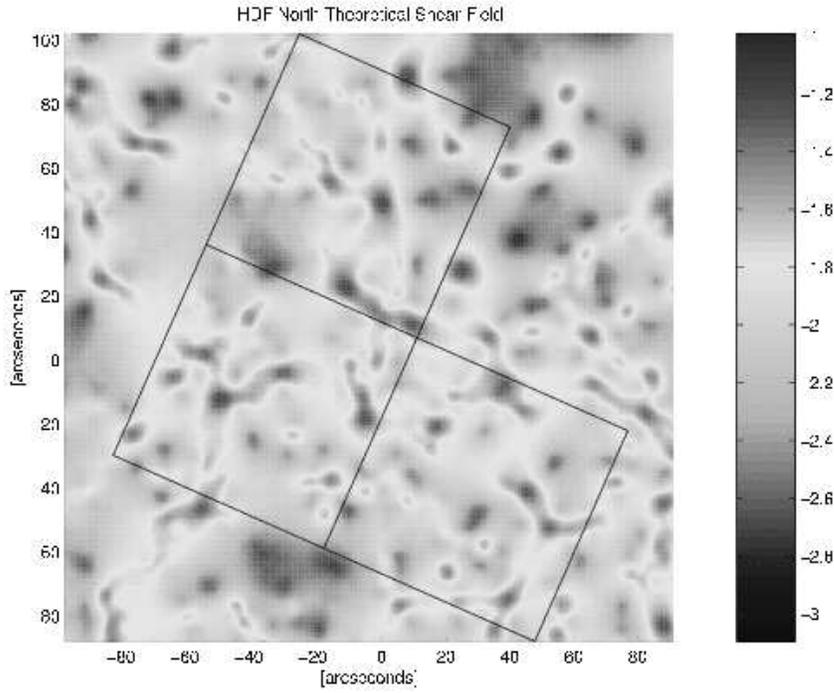}%
}
}
  \caption{Theoretical shear field in the region of the HDF--North.  See text
for model details.}
\end{figure}

Shown in Fig.\ 1 is the theoretical shear field in the region of the HDF--North
that would be produced by the 427 galaxies in Cohen et al.\ (2000) and
Cohen (2001) for which $R \le 23$, and both spectroscopic redshifts and
rest frame blue luminosities are known.  The median redshift of the lenses
is $\sim 0.6$.
Source galaxies were assumed to have apparent magnitudes in the
range $19 < I < 25$, and were distributed randomly on the sky.
A flat, $\Lambda$--dominated
cosmography with $H_0 = 70$~km/s/Mpc, $\Omega_0 = 0.3$, and $\Lambda_0 = 0.7$
was adopted for Fig.\ 1.  A fiducial $L^\ast$ galaxy model for which the
velocity dispersion of the halo is
$\sigma_v^\ast = 150$~km/s and the outer scale radius is $s^\ast = 100~h^{-1}$~kpc
was used and the source galaxies were assumed to follow a redshift distribution
of the form:
\begin{equation}
P(z|I) = \frac{\beta z^{2}\exp[-(z/z_{0})^{\beta}]}{\Gamma(3/\beta)z_{0}^{3}}
\label{zdist}
\end{equation}
(e.g., Baugh \& Efstathiou 1993),
which is in good agreement with the redshift surveys of
 LeF\`evre et al.\ (1996) and LeF\`evre et al.\ (2004).  Here
\begin{equation}
z_{0} = k_{z}[z_{m} + z_{m}'(I - I_{m})],
\end{equation}
where $z_m$ is the median redshift, $I_m$ is the median $I$-band
magnitude, and $z_m'$ is the derivative of the median redshift with respect
to $I$.  Extrapolating the results of LeF\`evre et al.\ (2004) 
to a sample of galaxies with
$19 < I < 25$, $z_m = 0.86$, $z_m' = 0.15$, $k_z = 0.8$,
and $\beta = 1.5$.   The median redshift of the sources is $\sim 1.2$.
Fig.\ 1 shows the mean over 6500 Monte Carlo realizations
of the weak lensing shear field and a
close comparison of this figure with an image of the HDF--North shows
that the peaks in theoretical shear field correspond to the brightest galaxies
in the HDF--North.

\section{Multiple Deflections}

This probability that a given source will have been
weakly--lensed by one or more foreground galaxies is, of course, a strong
function of the actual value of the shear, $\gamma$, due to a given
weak lensing deflection.  That is, it is
much more likely for a distant galaxy to be lensed by a foreground
galaxy which
produces an insignificant weak shear of
$\gamma \sim 10^{-6}$ than, say, a large weak
shear of $\gamma \sim 0.01$.  Therefore, in order to discuss the total number
of weak deflections that a given source galaxy is likely to encounter, a
decision has to be made as to what minimum value of $\gamma$ qualifies as a
``significant'' deflection.

A typical value for the net shear due to galaxy--galaxy lensing is
$\gamma \sim 0.005$ (see, e.g., the observational papers cited in the
Introduction) and this value of $\gamma$ was used as a baseline
for computing the number of weak lensing deflections that source galaxies
had undergone.  Specifically, the probability
that a given galaxy was lensed by $N_D$ foreground galaxies, $P(N_D)$,
where each {\it individual deflection} gave rise to
a shear of $\gamma > 0.005$ was computed.  
That is, $P(N_D = 2)$ is the probability that
a given galaxy had been lensed by two individual
foreground galaxies, each of which
lensed the distant galaxy at a level that is comparable to or greater
than the expected net shear due to galaxy--galaxy lensing.
Therefore,
the results shown here are an extremely conservative estimate
of the frequency of multiple deflections.

\begin{figure}
\centerline{
\scalebox{0.60}{%
\includegraphics{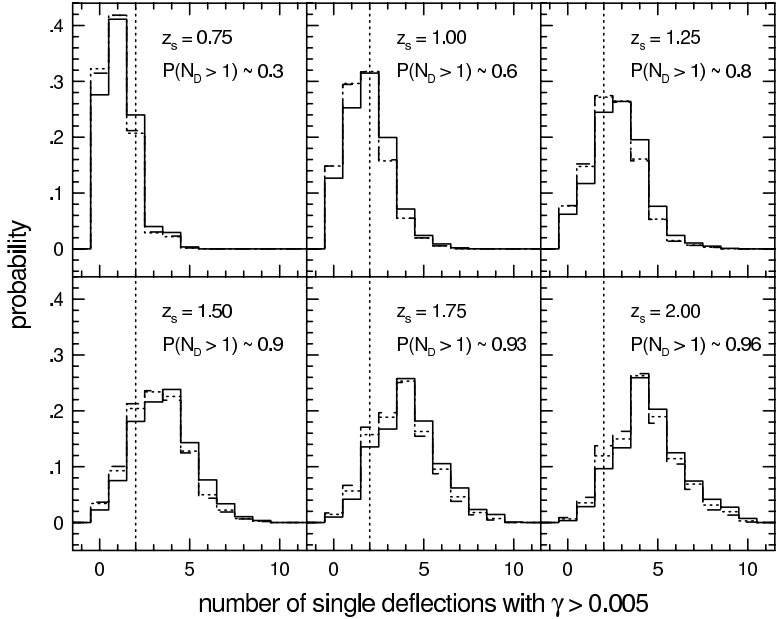}%
}
}
\vskip -6.7cm
  \caption{Probability of a given source galaxy in the 
HDF--North undergoing $N_D$ weak
deflections of individual magnitude $\gamma > 0.005$.  Panels correspond
to sources with different redshifts, $z_s$.  Solid histogram: flat,
$\Lambda$--dominated model.  Dashed histogram: open model.  Dotted
histogram: Einstein--de Sitter model.  Vertical dotted line shows
$N_D = 2$.
}
\end{figure}

Shown in Fig.\ 2 is the theoretical probability distribution
function, $P(N_D)$, for source galaxies in the HDF--North with a given
redshift, $z_s$.  The fiducial halo model and redshift distribution
adopted for Fig.\ 1 were also used here.  In addition, the cosmography
was varied from the flat, $\Lambda$--dominated model used in Fig.\ 1
to include both an open ($\Omega_0 = 0.3$, $\Lambda_0 = 0$) and an
Einstein--de Sitter model for Fig.\ 2.
From this figure, then, 
the probability that a source at redshift $z_s \sim 1$
has been lensed at a significant level by {\it two or more}
foreground galaxies is of order 60\%.  Also, it is clear from Fig.\ 2
that the probability of multiple deflections occurring is essentially
independent of the cosmography.
Instead, the frequency is influenced most strongly by the number of
massive lens galaxies that are close to the line of sight (i.e., 
galaxy--galaxy lensing provides information primarily about the potentials
of the lens galaxies, not cosmology per se; see also BBS).

At a source redshifts 
$z_s \sim 1.5$, the probability of a source galaxy encountering
multiple deflections of $\gamma > 0.005$ increases to of order
90\%.  Therefore, in a
deep data set for which
the median redshift is $\go 1$, it should be expected
that multiple weak deflections of a
substantial magnitude are very likely to have occurred. Of course, in the
case of individual deflections for which $\gamma < 0.005$, the probability
of comparable multiple deflections occurring at any given source redshift
will be greater than the results shown in Fig.\ 2.

\section{Effects of Halo Parameters on Multiple Deflections}

While the occurrence of multiple deflections is largely independent
of the cosmology, it is certainly not independent of the details of
the halo parameters.  In Figs.\ 1 and 2 a fiducial $L^\ast$ galaxy halo with
$\sigma_v^\ast = 150$~km/s and $s^\ast = 100$~kpc was adopted.  However,
galaxy--galaxy lensing constraints on the characteristic velocity
dispersion of the lens galaxies range from $\sigma_v^\ast \sim 135$~km/s (e.g., McKay
et al.\ 2001; Hoekstra, Yee, \& Gladders 2004) to $\sigma_v^\ast \sim 165$~km/s 
(e.g., Kleinheinrich
et al.\ 2004).  Constraints on the characteristic scale radius are
few, and based on the galaxy--galaxy lensing observations of
Hoekstra, Yee, \& Gladders (2004), the halos of $L^\ast$ galaxies may be as large
as $s^\ast \sim 185~h^{-1}$~kpc.

In this section, then, the effects of varying the halo parameters on
the occurrence of multiple deflections is investigated.  
Since the occurrence of multiple deflections is only weakly
dependent on the cosmography, for the remainder of this paper
only the flat, $\Lambda$--dominated
model was used.  Again, all source galaxies were assumed to have
magnitudes in the range $19 < I < 25$, with redshifts determined by
eqns.\ (2.4) and (2.5) above.  
The halo parameters of the lens galaxies in the HDF--North
and flanking fields were then varied as follows:
$\sigma_v^\ast = 135$~km/s, 150~km/s, and 165~km/s; $s^\ast = 50~h^{-1}$~kpc,
$100~h^{-1}$~kpc, and $200~h^{-1}$~kpc.

\begin{figure}
\centerline{
\scalebox{0.60}{%
\includegraphics{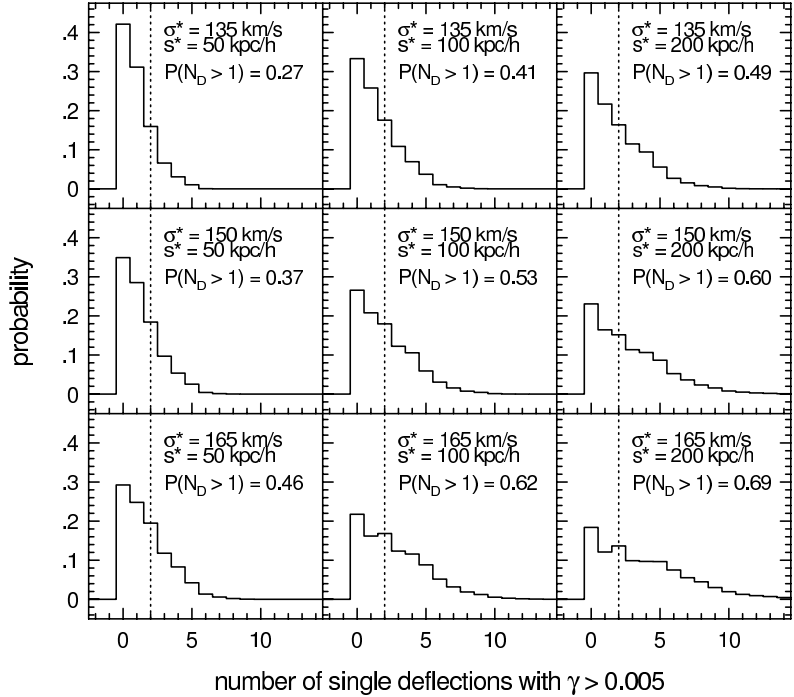}%
}
}
\vskip -5.7cm
  \caption{Probability of a given source galaxy in the 
HDF--North undergoing $N_D$ weak
deflections of individual magnitude $\gamma > 0.005$.  Panels correspond
to different halo models, indicated by the values of $\sigma_v^\ast$ and
$s^\ast$.  In all cases a $\Lambda$CDM model was used.
Unlike Fig.\ 2, here $P(N_D)$ was computed over the entire
redshift distribution of sources with $19 < I < 25$.  
Vertical dotted line shows
$N_D = 2$.
}
\end{figure}

Shown in Fig.\ 3 is $P(N_D)$ for all source galaxies in the HDF--North
with $19 < I < 23$.  Like Fig.\ 2, the minimum value for a ``deflection''
to be counted in this figure is $\gamma = 0.005$.  Unlike Fig.\ 2, however,
$P(N_D)$ has been computed over the entire redshift distribution of the
sources.  From this figure, then, the probability that a given source
in the HDF--North
has been lensed more than once is 27\% for the lowest mass $L^\ast$ halo
(upper left panel), 53\% for the fiducial halo of Figs.\ 1 and 2 (center
panel), and 69\% for the highest mass $L^\ast$ halo (bottom right panel).
Therefore, it is clear that the mass adopted for the halo of an $L^\ast$
galaxy has a rather substantial effect on the number of multiple deflections
that occur in a galaxy--galaxy lensing calculation.

\begin{figure}
\centerline{
\scalebox{0.60}{%
\includegraphics{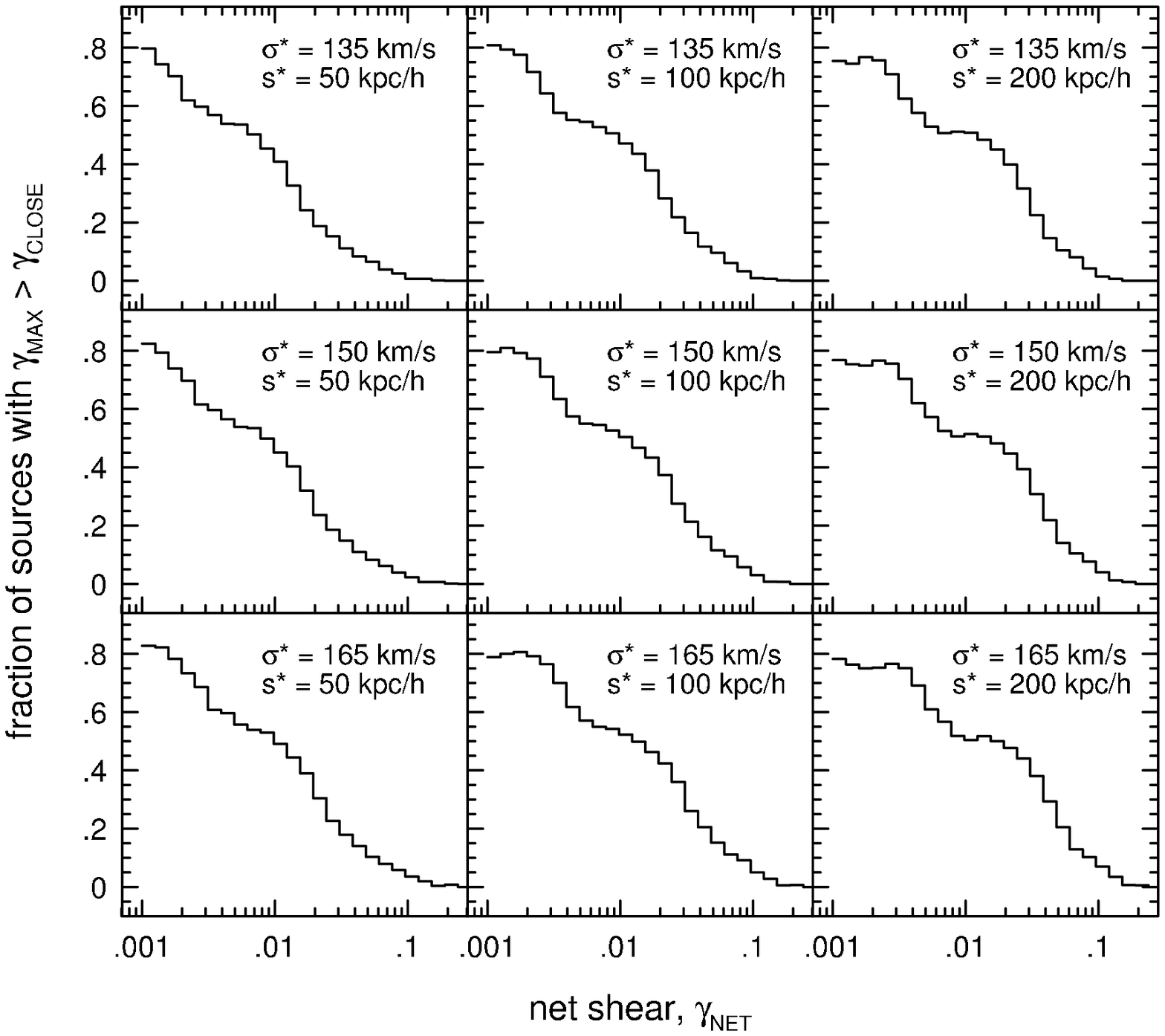}%
}
}
\vskip -5.7cm
  \caption{Fraction of sources for which the maximum value of a single
deflection, $\gamma_{\rm max}$, exceeds the value of the shear due
to the closest lens on the sky, $\gamma_{\rm close}$.
All sources with $19 < I < 25$ were included in the calculations.
}
\end{figure}

Since the halos are roughly isothermal and, therefore, the shear decreases
with projected radius approximately as $\gamma(\theta) \propto \theta^{-1}$, it
is interesting to ask whether the closest lens to a given source (in
projection on the sky) is necessarily the strongest lens.  That is, for
a source that undergoes multiple deflections, is the strongest deflection
most likely to come from the nearest lens on the sky?  The answer to
this question is ``No,'' and is shown clearly by Fig.\ 4.  In the case
of sources for which the net shear (i.e., the total shear after all
weak deflections have occurred) is $\gamma \lo 0.01$, more than 50\% of the time
the strongest lens is not the closest lens on the sky.  Note that this result
is essentially independent of the characteristic halo parameters that were
chosen, which is consistent with the fact that, by and large, it is only
the most massive galaxies that contribute substantially to the overall
galaxy--galaxy lensing signal.

Figs.\ 5 and 6 demonstrate a result that is perhaps somewhat 
counter--intuitive: multiple weak lensing deflections, on average, give
rise to a greater net shear on the source galaxies and a greater mean
tangential shear about the lens centers.  That is, multiple deflections
do not simply cancel one another out, leading to little or no net
shear on the source galaxies.  Fig.\ 5 shows probabilities for the
distribution of the ratio of the maximum value of any given single deflection, 
$\gamma_{\rm max}$, to that of the net shear, $\gamma_{\rm net}$, for
all sources with $19 < I < 25$ that underwent more than one
deflection of any magnitude (i.e., $\gamma > 0$).  
The vertical dotted line shows the median
value of the distribution, and for all halo models the median is less
than 1.  That is, more than 50\% of the time, the net shear on a given
source is greater than the maximum single deflection that it underwent.

\begin{figure}
\centerline{
\scalebox{0.60}{%
\includegraphics{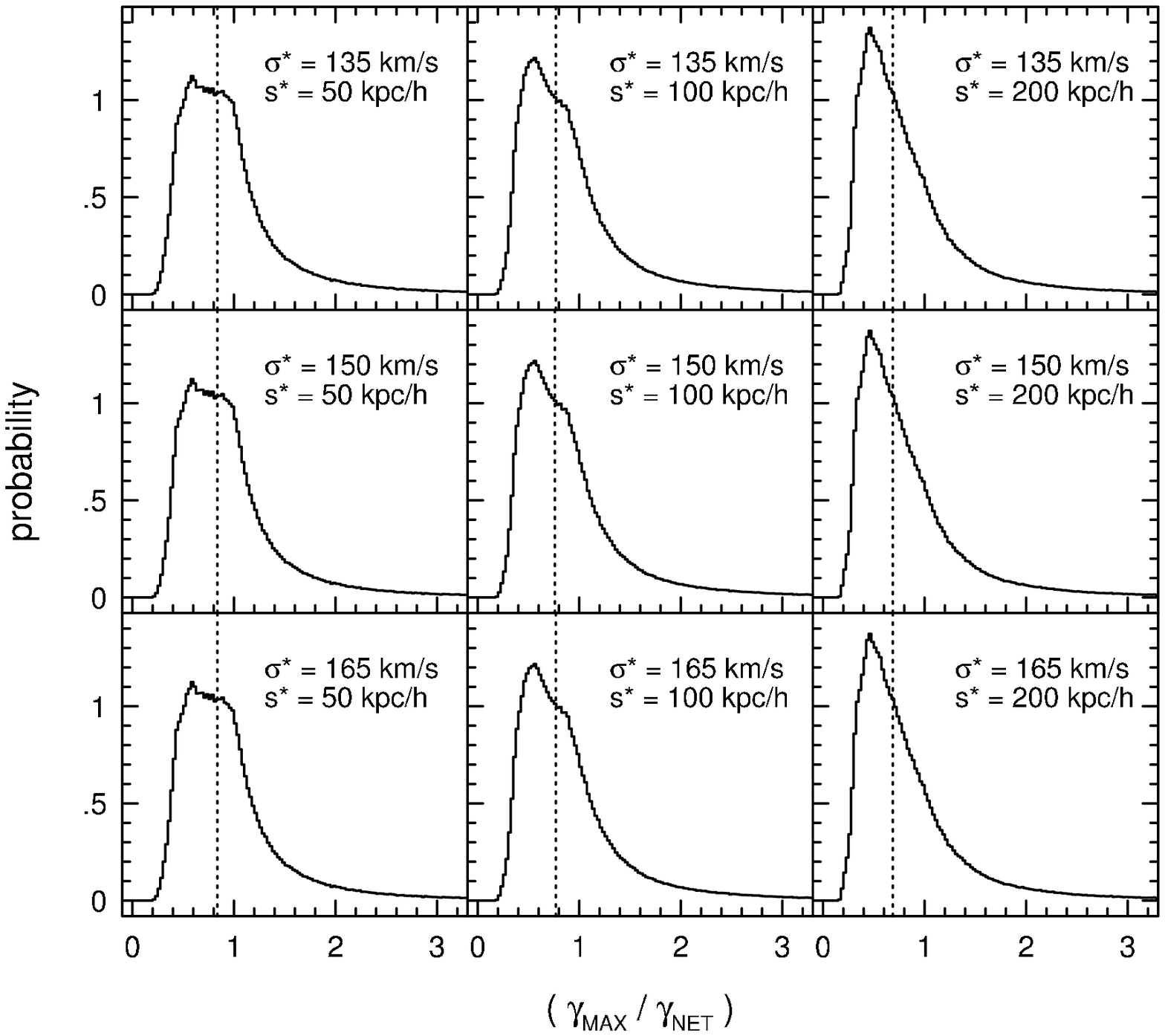}%
}
}
\vskip -5.7cm
  \caption{Probability distribution for the ratio of the maximum shear
due to a single deflection, $\gamma_{\rm max}$, to that of the 
the net shear, $\gamma_{\rm net}$.  Vertical dotted line shows the
median of the distribution. 
All sources with $19 < I < 25$ that underwent more than one
deflection of magnitude $\gamma > 0$ were included in the calculations.
}
\end{figure}

Fig.\ 6 shows the mean tangential shear, computed about the lens 
centers in the HDF--North.  Squares show the results of the proper
inclusion of multiple deflections for all sources, and crosses show
the result of lensing each source solely by the closest lens on the
sky (i.e., a ``single deflection'' calculation).  
In all cases, the inclusion of multiple deflections gives rise
to a larger mean tangential shear on scales $\theta \go 1''$, and in the 
case of the more massive halos, the increase in the shear is quite
substantial.  
This figure, then emphasizes the need for a correct, multiple deflection
calculation when using observations of the mean tangential shear to
constrain the halo properties of the lens galaxies (i.e., the 
comparison of single
deflection calculations to observations leads to an inferred halo
mass for $L^\ast$ galaxies that is too large).

\begin{figure}
\centerline{
\scalebox{0.60}{%
\includegraphics{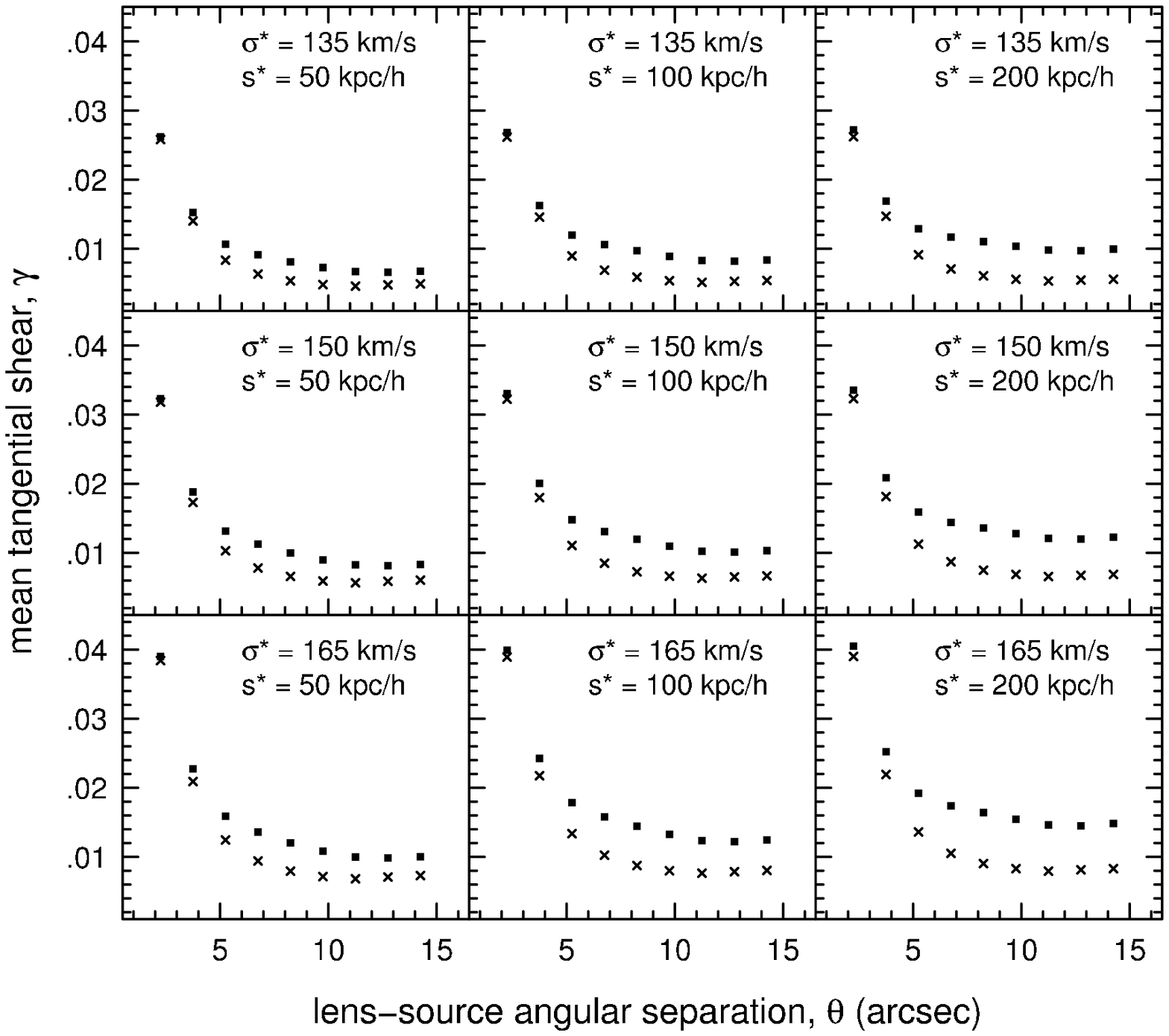}%
}
}
\vskip -5.7cm
  \caption{Mean tangential shear computed about the lenses in the
HDF--North.  Squares show the results of the multiple deflection
calculations, crosses show the result of lensing each source galaxy
solely by the nearest lens on the sky.
All sources with $19 < I < 25$ 
were included in the calculations.
}
\end{figure}

\section{Correlated Image Ellipticities and Cosmic Shear}

In addition to giving rise to a generally larger net shear on source
galaxies and a larger mean tangential shear about the lens centers,
multiple deflections in galaxy--galaxy lensing give rise to correlated
ellipticities in the images of the galaxies.  This is, of course,
precisely the effect of cosmic shear, but in the case of 
galaxy--galaxy lensing, this is merely
the very small $k$ end of the power spectrum (i.e., the highly non--linear
regime) that is contributing to the total cosmic shear signal (i.e., 
when computed over all structures along the line of sight).

\begin{figure}
\centerline{
\scalebox{0.60}{%
\includegraphics{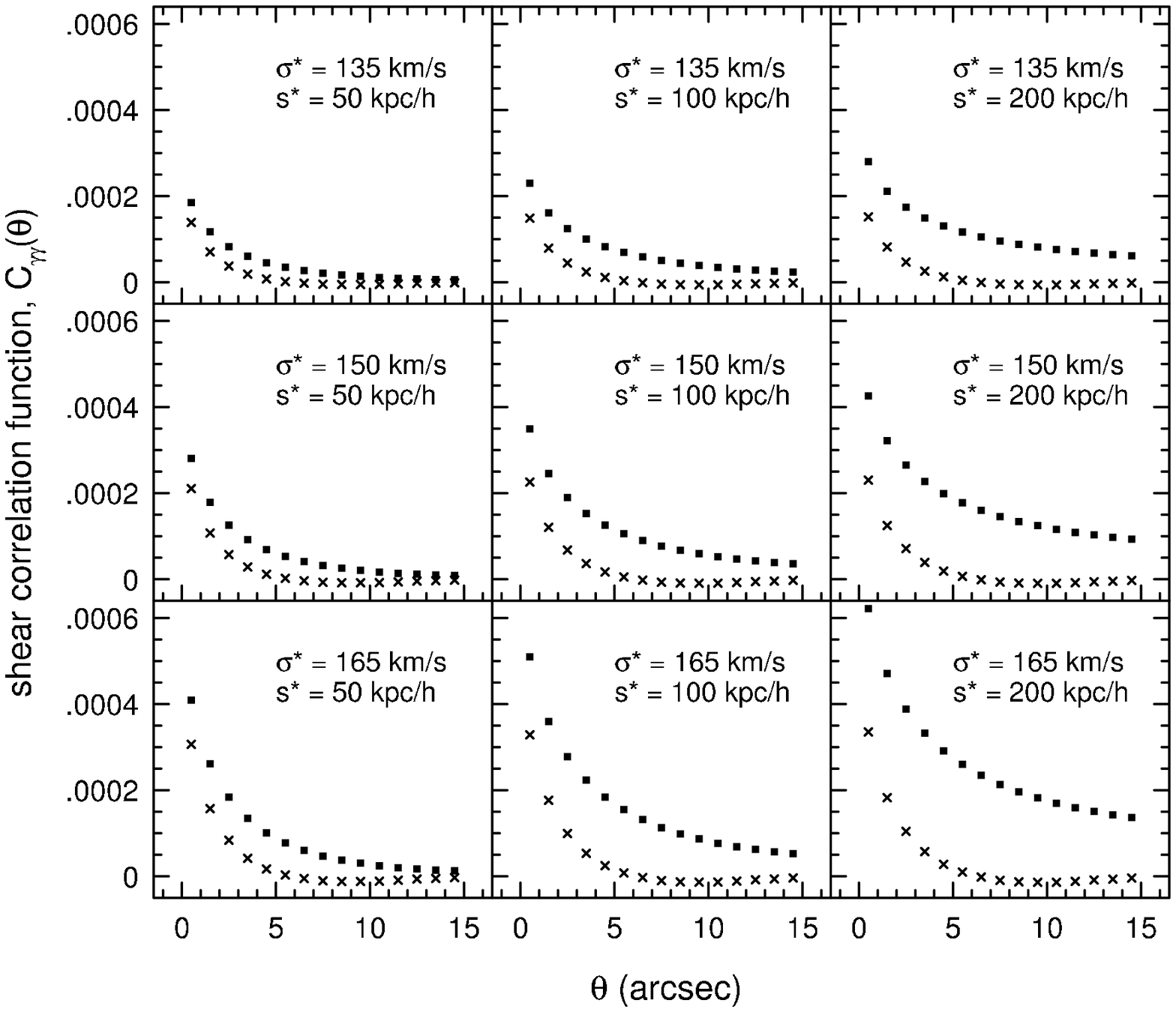}%
}
}
\vskip -5.7cm
  \caption{Shear correlation function, $C_{\gamma\gamma}(\theta)$, as
a function of angular scale and halo parameters.
Squares show the results of the multiple deflection
calculations, crosses show the result of lensing each source galaxy
solely by the nearest lens on the sky.
All sources with $19 < I < 25$ 
were included in the calculations.
}
\end{figure}

The degree to which multiple deflections give rise to correlated image
ellipticities and, hence, contribute to the cosmic shear signal is
a strong function of the mass adopted for the halo of an $L^\ast$ galaxy.
This is shown in Fig.\ 7, where the shear correlation function,
\begin{equation}
C_{\gamma\gamma}(\theta) \equiv \left< \vec{\gamma}_i \cdot \vec{\gamma}_j^\ast
\right> ,
\;\;\;\;\; i \ne j
\end{equation}
is shown as a function of the halo parameters.  
The mean value is computed for all foreground--background pairs of
galaxies separated by angles $\theta \pm \delta\theta/2$
on the sky (see, e.g.,
Blandford et al.\ 1991).
Here $\vec{\gamma}_{i}$ is the image shape of galaxy $i$  and
$\vec{\gamma}_{j}^\ast$ is the complex conjugate
of the image shape of galaxy $j$.  The image shape is defined as
\begin{equation}
\gamma \equiv \frac{a^2-b^2}{a^2+b^2} e^{2i\phi},
\end{equation}
where $a$ and $b$ are the major and minor axes of the image equivalent
ellipse and $\phi$ is its position angle.
As in Fig.\ 6, squares
show the results of the full multiple deflection calculations and crosses
show the results of calculations in which each source was lensed solely
by the closest lens on the sky.  From this figure, then, the single
deflection calculations do not give rise to correlated image ellipticities
on scales $\theta \go 5''$ (on smaller scales, of course, the images of distant
sources that have been lensed by the identical foreground galaxy will be
correlated because of the tangential alignment about the lens center).
In the multiple deflection calculations, however, sufficiently massive
halos give rise to correlated ellipticities that persist to significantly
large angles, just due to the galaxy--galaxy lensing signal alone.  

Oftentimes the phrase ``cosmic shear'' is interpreted to mean ``lensing
by large--scale structure'', and this is true on large angular 
scales where it is only structure in the linear regime that is responsible
for the gravitational lensing.  Properly, however, cosmic shear is the
lensing of distant galaxies by {\it all} the mass along the line of sight,
including highly non--linear structures.  This is why, on small angular
scales, it is necessary to use large simulations (e.g., Jain, Seljak, \&
White 2000; Valageas, Barber, \& Munshi 2004; Vale \& White 2003) to 
make accurate theoretical predictions of cosmic shear.

From Fig.\ 7, it is clear that, depending upon the characteristic mass
of the halos of $L^\ast$ galaxies, galaxy--galaxy lensing will contribute
to the cosmic shear signal as measured, for example, via the 
top hat shear variance,
\begin{equation}
\left< \gamma^2 \right> = \frac{2}{\pi \theta^2}\int_0^\infty \frac{dk}{k}~
P_\kappa(k) \left[ J_1 (k\theta) \right]^2.
\end{equation}
Here  $P_\kappa$ is the power spectrum of the projected mass density
of the universe, $J_1$ is a Bessel function of the first kind, and
$\theta$ is the size of the aperture over which the mean is computed.
In an observational data set, the function is computed as
\begin{equation}
\left< \gamma^2 \right> = 
\frac{1}{N (N-1)} \sum_{i\ne j} \gamma_i \cdot \gamma_j^\ast,
\end{equation}
for all galaxies within an aperture of size $\theta$ on the sky
(see, e.g., H\"ammerle et al.\ 2002).  The above published predictions for
$\left< \gamma^2 \right>$ in CDM models have been based on simulations
in which the mass and force resolution were not quite adequate to
resolve the halos of $L^\ast$ galaxies particularly well and, so, it is
difficult to be certain how accurate the theoretical predictions truly
are on very small angular scales.

\begin{figure}
\centerline{
\scalebox{0.65}{%
\includegraphics{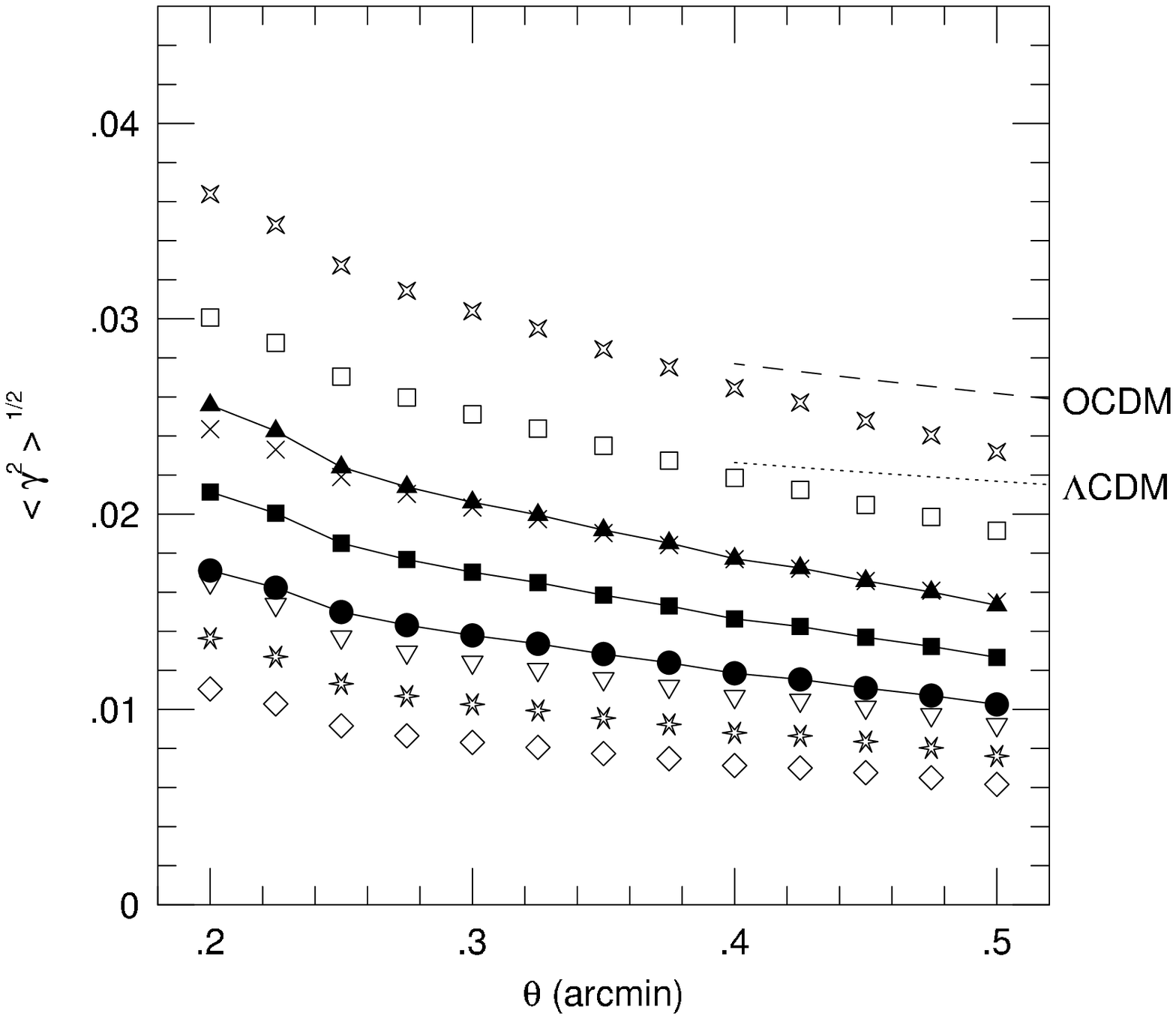}%
}
}
\vskip -6.5cm
  \caption{The r.m.s.\ cosmic shear as a function of aperture size, $\theta$.
Points show the results from ray--tracing simulations of galaxy--galaxy 
lensing in the HDF--North for which
all sources with $19 < I < 25$ 
were included in the calculations.  From bottom to top, the points
correspond to increasing halo mass (see text).  Filled squares show results
for a fiducial $L^\ast$ galaxy halo for which $\sigma_v^\ast = 150$~km/s
and $s^\ast = 100~h^{-1}$~kpc. Filled circles and filled triangles show
results for a $\sim 20$\% 
decrease in the fiducial halo mass and a $\sim 20$\% increase
in the fiducial halo mass, respectively.  Dashed and dotted lines show
theoretical predictions for open CDM and $\Lambda$CDM, respectively, 
based on calculations from Jain, Seljak, \& White (2000). 
}
\end{figure}

The symbols in Fig.\ 8 show the r.m.s.\ cosmic shear, 
$\left< \gamma^2 \right>^{1/2}$, computed within circular apertures
of radius $\theta$ in the full multiple deflection calculations
of galaxy--galaxy lensing
in the HDF--North.  That is, the symbols in this figure show the r.m.s.\
cosmic shear {\it due to galaxies alone}, without any contribution from
linear or quasi--linear structures along the line of sight.  From
bottom to top, the point types correspond to increasing the halo mass from
a minimum mass of
$M^\ast = 0.83 \times 10^{12} \Msun$ (open diamonds) to a
 maximum mass of $M^\ast = 4.05\times
10^{12} \Msun$ (four--pointed stars).  The mass of the fiducial halo with 
$\sigma_v^\ast = 150$~km/s and $s^\ast = 100~h^{-1}$~kpc is
$M^\ast = 1.67\times 10^{12} \Msun$, and the r.m.s.\ cosmic shear in this
case is shown by the solid squares.  Also shown is the prediction for
$\left< \gamma^2 \right>^{1/2}$ for $\Lambda$CDM and open CDM from 
the simulations of Jain, Seljak, \& White (2000) for a source galaxy population
with median redshift of $z_s \sim 1.2$ (i.e., similar to the HDF--North
raytracing simulations for sources with $19 < I < 25$).  The $\Lambda$CDM
and OCDM predictions do not extend below $\theta \sim 0.4'$.

The filled circles and filled triangles show the r.m.s.\
cosmic shear produced by galaxy--galaxy lensing alone for cases in
which the mass of the halo of an $L^\ast$ galaxy is $\sim 20$\% less
than the fiducial halo mass and $\sim 20$\% greater than the fiducial
halo mass, respectively.  At $\theta \sim 0.5'$, then, a change in the
fiducial halo mass of only 20\% results in a change in the predicted
cosmic shear signal that is quite comparable to the predicted differences
between two rather different cosmological models.  In other words, if
one wishes to use observations of cosmic shear on small angular 
scales to constrain the cosmography (through, e.g., a comparison to
ray--tracing simulations) it is vital that the simulations have followed
the formation of the growth of non--linear structures very accurately.

The r.m.s.\ cosmic shear due solely to galaxy--galaxy lensing extrapolates
to zero at $\theta \sim 1'$ for the fiducial halo model with $\sigma_v^\ast
= 150$~km/s and $s^\ast = 100~h^{-1}$~kpc.  It is, therefore, only on
scales $\lo 1'$ that contributions of galaxy--galaxy lensing to the
cosmic shear signal are likely to be of importance.  

\section{Conclusions}

The occurrence and effects of multiple weak deflections due to galaxy--galaxy
lensing were investigated for a deep data set in which $z_{\rm lens} \sim 0.6$
and $z_{\rm source} \sim 1.2$.  Ray--tracing simulations of galaxy--galaxy
lensing by lens galaxies with $R \le 23$ in the Hubble Deep Field (North)
and flanking fields
were used to compute the net shear on source galaxies with magnitudes
in the range $19 < I < 25$.
Both the redshifts and the rest frame blue luminosities
of the lenses are known, which allows for a detailed theoretical prediction
of galaxy--galaxy lensing in the HDF--North, given a particular cosmography
and a model for the halos of the lens galaxies.  The primary conclusions
from this work are:

\begin{itemize}
\medskip
\item Multiple weak deflections are commonplace in such deep data sets,
and the proper inclusion of multiple deflections is important to a 
correct prediction for the net shear experienced by the majority of
source galaxies.
\medskip
\item For a given source redshift, the probability
of multiple weak deflections is largely insensitive to the cosmography
(i.e., galaxy--galaxy lensing is much more sensitive to the details
of the gravitational potentials of the lens galaxy halos than it is
to the values of the cosmological parameters).
\medskip
\item Compared to a single deflection calculation in which sources are
lensed solely by the nearest lens on the sky, a full multiple deflection
calculation leads both to a larger net shear for most individual sources and a
larger tangential shear about the lens centers.
\medskip
\item Multiple weak deflections give rise to correlated image ellipticities
(i.e., ``cosmic shear'' due to power on highly non--linear scales).
\medskip
\item On angular scales $\theta \lo 1'$, galaxy--galaxy lensing alone accounts for
a substantial amount of the cosmic shear signal expected in $\Lambda$CDM
and open CDM models.   The magnitude of the signal
is, however, very sensitive to the characteristic
halo mass and, hence, accurate comparisons of observations of cosmic shear
and theory on such scales rely heavily on the ability of simulations to
follow the growth of the non--linear power spectrum with high accuracy. 
\end{itemize}

\begin{acknowledgments}
It is a great pleasure to thank Yannick and Georges for the invitation
to come to the beautiful city of Lausanne, and for their tremendous
efforts to insure that the Symposium was a true success.
Support under NSF contract AST-0098572 is gratefully acknowledged.
\end{acknowledgments}

\end{document}